\documentclass{PoS}
\usepackage{latexsym,amssymb,amsfonts,amsmath}

\newcommand{\nyp}[3]{{\bf #1} (#3) #2}
\newcommand{\JHEP}{JHEP }

\newcommand{\diag}{{\rm diag}}
\newcommand{\tr}{{\rm Tr}}
\newcommand{\la}{\langle}
\newcommand{\ra}{\rangle}

\renewcommand{\Re}{{\rm Re\,}}

\newcommand{\f}[2]{\frac{#1}{#2}}

\renewcommand{\Re}{{\rm Re}}

 \newcommand{\tf}[2]{{\textstyle\f{#1}{#2}}}
\newcommand{\tdU}{{U}^{\rm (td)}}

\title{Chiral transition, eigenmode localisation and Anderson-like models}

\ShortTitle{Chiral transition, eigenmode localisation and Anderson-like models}

\author{\speaker{Matteo Giordano} \thanks{Supported by OTKA under the
    grant OTKA-K-113034.}\\
        Institute for Theoretical Physics, E\"otv\"os University,\\
        and MTA-ELTE Lattice Gauge Theory Research Group,\\
        P\'azm\'any P. s\'et\'any 1/A, H-1117 Budapest, Hungary\\
        E-mail: \email{giordano@bodri.elte.hu}}

\author{Tam\'as G.\ Kov\'acs\protect{\footnotemark[2]}\mbox{ }
  \thanks{Supported by the Hungarian Academy of Sciences under
    ``Lend\"ulet'' grant No. LP2011-011.} \\
       Institute for Nuclear Research of the Hungarian Academy of Sciences, \\
       Bem t\'er 18/c, H-4026 Debrecen, Hungary 
       \\
       E-mail: \email{kgt@atomki.mta.hu}}
\author{Ferenc Pittler\\
       HISKP(Theory), University of Bonn,\\
       Nussallee 14-16, D-53115,Bonn, Germany\\
       E-mail: \email{pittler@hiskp.uni-bonn.de}}

\abstract{We discuss chiral symmetry restoration and eigenmode
  localisation in finite-temperature QCD by looking at the lattice
  Dirac operator as a random Hamiltonian. We argue that the
  features of QCD relevant to both phenomena are the presence of order
  in the Polyakov line configuration, and the correlations that this
  induces between spatial links across time slices. This ties the fate
  of chiral symmetry and of localisation of the lowest Dirac
  eigenmodes to the confining properties of the theory. We then show
  numerical results obtained in a QCD-inspired Anderson-like toy
  model, derived by radically simplifying the QCD dynamics while
  keeping the important features mentioned above. The toy model
  reproduces all the important qualitative aspects of chiral symmetry
  breaking and localisation in QCD, thus supporting the central role
  played by the confinement/deconfinement transition in triggering
  both phenomena.}
 
\FullConference{34th annual International Symposium on Lattice Field
  Theory,\\ 
  24-30 July, 2016\\
  University of Southampton, UK}

\begin{document}

\section{Introduction}
\label{sec:intro}

As is well known, in QCD both the confining and chiral properties of
the theory undergo a rapid change around a temperature $T_c\simeq 155~
{\rm MeV}$~\cite{Borsanyi:2010cj}. Moreover, in models where there is
a genuine phase transition, rather than just an analytic crossover
like in QCD, deconfinement and the chiral transition take place at the 
same temperature. Such models include SU$(2)$ and SU$(3)$ pure gauge
theory, SU$(3)$ gauge theory with uninmproved staggered fermions on
$N_T=4$ lattices~\cite{unimproved}, and SU$(3)$ gauge theory with
adjoint fermions~\cite{adjoint}.\footnote{In the latter case the
  deconfinement transition is accompanied by a first-order chiral
  transition signalled by a jump in the chiral condensate, while a
  second transition, fully restoring chiral symmetry, takes place at a
  higher temperature.}  
A convincing explanation of why deconfinement and the chiral transition
take place together is still lacking.

In recent years there has been growing evidence of a third phenomenon
taking place in QCD around $T_c$, namely the change in the
localisation properties of the low-lying eigenmodes of the Dirac
operator~\cite{GGO2,KGT,KP,BKS,KP2,crit,UGPKV,Cossu:2016scb}. While
below $T_c$ all the modes are delocalised~\cite{VWrev}, above 
$T_c$ the lowest modes are spatially localised on the scale of the
inverse temperature, up to a critical, temperature-dependent point
$\lambda_c(T)$ in the spectrum (``mobility edge''); modes above
$\lambda_c$ are again delocalised. The localisation/delocalisation
transition at the mobility edge is a second-order phase transition
with divergent correlation length~\cite{crit}. Extrapolating
$\lambda_c(T)$ as $T$ is decreased, one finds that $\lambda_c(T)$
vanishes at a temperature compatible with $T_c$. The low-lying modes
play a crucial role in physical observables, in particular in the
formation of a chiral condensate, and therefore for the spontaneous
breaking of chiral symmetry (S$\chi$SB). It is therefore not so
surprising that the localisation properties of these modes change at
the chiral transition. 

The appearance of localised modes in the deconfined/chirally restored
phase is not a unique feature of QCD, but has been observed also in other
QCD-like models~\cite{GGO2,KP,GKKP}. In models where the transition is
sharp one can ask more meaningfully whether localisation appears right
at the critical point. In the above-mentioned model with unimproved
staggered fermions~\cite{unimproved}, which displays a first-order
deconfining and chiral transition, the onset of localisation has been 
shown to take place precisely at the critical temperature~\cite{GKKP}. 
Understanding localisation might therefore help in understanding the
relation between deconfinement and the chiral transition.

\section{QCD, the Anderson Model and the Dirac-Anderson Hamiltonian}  
\label{sec:DAH}

The best known model displaying localisation of eigenmodes is the
celebrated Anderson model (AM)~\cite{Anderson58}, which provides an
approximate description of (non-interacting) electrons in ``dirty''
conductors. The effect of impurities is modelled by a random on-site
potential, added to the usual hopping terms of the tight-binding
Hamiltonian; the amount of impurities/disorder is controlled by
the width $W$ of the range in which the random potential takes its
values. In the AM, eigenmodes at the band edge are localised, for
energies beyond the so-called mobility edge, and the
localisation/delocalisation transition within the spectrum is a
second-order phase transition with divergent correlation
length. Concerning localisation of the eigenmodes, QCD and the AM are 
thus completely analogous. Since in the AM the position of the
mobility edge is determined by $W$, while in QCD it is controlled 
by the temperature $T$, one is led to identify $T$ as the parameter
effectively controlling the amount of disorder in QCD. 
There is, however, more than a simple analogy between the two models. 
In fact, the correlation-length critical exponent in QCD was found to
be compatible~\cite{crit} with that of the 3D unitary
AM~\cite{nu_unitary}.\footnote{The unitary AM is obtained by
  multiplying the hopping terms in the AM by random phase factors,
  mimicking the presence of a magnetic field.} 
Here ``unitary'' refers to the symmetry class of the model in the
classification of Random Matrix Theory (RMT)~\cite{VWrev}. Moreover,
eigenmodes at the mobility edge are multifractal~\cite{UGPKV}, with  
multifractal exponents matching those of the 3D unitary AM~\cite{UV}. 
These results have been obtained using the staggered discretisation of
the Dirac operator.

The fact that the transitions in the two models share the same
universality class calls for an explanation. From the point of view of
RMT, the QCD staggered Dirac operator is a random matrix of the
unitary class, with off-diagonal noise provided by the fluctuations of 
the gauge links. Moreover, QCD at finite temperature is a 4D
model. The unitary AM, on the other hand, is 3D with mostly diagonal disorder.
A qualitative explanation for the matching of universality classes is
the following~\cite{BKS,GKP,GKP2}. For $T>T_c$ time-slices are
strongly correlated, QCD is effectively 3D, and the quark
eigenfunctions $\psi(t,{\vec x})$ are expected to look qualitatively
the same on all the time-slices. Moreover, working in the temporal gauge,
one sees that $\psi(t,{\vec x})$ obeys effective boundary conditions
in the temporal direction, which involve the local Polyakov line (PL),
$P(\vec x)$, namely $  \psi(N_T,{\vec x})=-P({\vec x})\psi(0,{\vec x})$. 
Here $N_T$ is the temporal extension of the system in lattice units.
Since $P(\vec x)$ fluctuates in space, it provides an effective 3D
source of diagonal disorder. This explanation leads, however, to another
question. The effective boundary conditions apply both above and below
$T_c$, and similarly the PL fluctuates in space in both phases of the
theory. One thus has to justify why QCD is effectively 4D at low $T$,
and why the effective boundary conditions are ineffective there. 

In order to better understand the relation between QCD and the unitary
AM, following Ref. \cite{GKP2}, we make the connection explicit
between the staggered Dirac operator $D_{\rm stag}$ and Anderson-type
Hamiltonians. We start from the ``Hamiltonian'' $H= -iD_{\rm stag}$,
split it into a ``free'' and an ``interaction'' part, $H=H_0+H_I$, and
then work in the basis of the ``unperturbed'' eigenvectors of $H_0$. 
The physically most sensible choice is to identify $H_0$ with the
temporal hoppings, $(H_0)_{xx'} = \tf{\eta_4(\vec x)}{2i}
[U_4(t,\vec x)\delta_{t+1,t'} - U_{-4}(t,\vec x)\delta_{t-1,t'}]
\delta_{\vec x,\vec x{}'}$, thus leaving the spatial hoppings as the
spatially isotropic interaction part, $(H_I)_{xx'} = 
\textstyle\sum_{j=1}^3 \tf{\eta_j(\vec x)}{2i}
[U_j(t,\vec x)\delta_{\vec x+\hat\jmath,\vec x{}'} - 
U_{-j}(t,x)\delta_{\vec x-\hat\jmath,\vec x{}'}]\delta_{t,t'}$.
Here $x=(t,\vec x)$, $U_\mu(x)=U_\mu(t,\vec x)$ are SU$(N_c)$
gauge links with $U_{-\mu}(x)=U^\dag_\mu(x-\hat\mu)$, and
$\eta_\mu(\vec x)$ are the usual staggered phases. In the temporal
diagonal gauge one has $U_4=\mathbf{1}$ $\forall t\neq N_T-1$, and
$U_4(N_T-1,\vec x)=P(\vec x)=\diag(e^{i\phi_a(\vec x)})$,
$a=1,\ldots,N_c$, having diagonalised each PL with a
time-independent gauge transformation. We choose $\sum_a \phi_a(\vec
x)=0$, any other choice leading to a unitarily equivalent Hamiltonian. 
In this gauge it is straightforward to diagonalise $H_0$, finding for
the unperturbed eigenvalues 
\begin{equation}
  \label{eq:evs}
  \lambda_0^{\vec x\,a\,k} = \eta_4(\vec x) \sin \omega_{ak}(\vec
  x)\,, \quad    \omega_{ak}(\vec x) = {\textstyle\f{1}{N_T}}( \pi
  + \phi_a(\vec x) + 2\pi k)\,,
\end{equation}  
where $\omega_{ak}(\vec x)$ are the effective Matsubara frequencies. 
The various indices refer to the spatial site $\vec x$ where the
eigenmode is localised, and the colour component $a$ and the temporal
momentum (TM) $k=0,\ldots,N_T-1$ to which it corresponds. In the basis
of the unperturbed eigenvectors one finds 
\begin{equation}
  \label{eq:d_and}
  \begin{aligned}
       { H}_{\vec x,\vec y} &=\textstyle
    \delta_{\vec x, \vec y} D(\vec x)
    + \sum_{j=1}^3\f{\eta_j(\vec x)}{2i}\left[
      \delta_{\vec x+\hat\jmath,\vec y} V_{+j}(\vec x)
      -\delta_{\vec x-\hat\jmath,\vec y} V_{-j}(\vec x)
    \right]\,,\\
    \left[D(\vec x)\right]_{ak,bl} &=
    \lambda_0^{\vec x\,a\,k}\delta_{ab}\delta_{kl}\,, \\
    \left[V_{\pm j}(\vec x)\right]_{ak,bl} &
    =\textstyle 
    \f{1}{N_T} \sum_{t=0}^{N_T-1}
    e^{i\f{2\pi t}{N_T}(l-k)} 
    e^{i\f{t}{N_T}[\phi_{b}(\vec x\pm \hat \jmath)-\phi_{a}(\vec
      x)]} 
    \left[\tdU_{\pm j}(t,\vec x)\right]_{ab}\,,
  \end{aligned}
\end{equation}
with $\tdU_{\pm j}(t,\vec x)$ the spatial gauge links in the
temporal diagonal gauge.  

The ``Dirac-Anderson Hamiltonian'' of Eq.~\eqref{eq:d_and} 
is a 3D Anderson-type Hamiltonian with internal degrees of freedom,
corresponding to colour and TM, and including both diagonal and
off-diagonal noise. Unlike the AM, here the control parameter (the
temperature) does not change the overall strength of the disorder
(which is always bounded in magnitude), but rather its type and its
spatial distribution, which are completely different in the confined
and deconfined phases of QCD. 

In the deconfined phase, the PL gets ordered along $\mathbf{1}$, with
fluctuations away from the ordered value forming ``islands'' of
``wrong'' PLs. This directly reduces the density of small unperturbed 
eigenvalues. Moreover, it leads to a reduction of the hopping terms'
entries that are off-diagonal in TM, as a consequence of the
correlations among spatial links on different time-slices induced by
the ordering. Since ``wrong'' PLs yield smaller unperturbed
eigenvalues, one expects them to provide an ``energetically''
favourable place for the quark eigenfunctions to live on, thus
allowing for smaller eigenvalues. Moreover, the mixing of different TM 
components is harder outside of these islands. In other words, the
``islands'' provide localising ``traps'' for the low eigenmodes. 
The approximate decoupling of different TM components leads to having
in practice $N_T$ weakly coupled, effectively 3D systems. The ordering
of the PLs is also expected to affect the spectral density of the low
modes, and therefore the fate of chiral symmetry. Indeed, mixing of TM
components tends to ``push'' the modes towards the
origin,\footnote{This can be seen qualitatively using lowest-order 
  perturbation theory.} and its reduction, combined with a smaller
density of low unperturbed modes, leads us to expect a vanishing
spectral density at zero, $\rho(0)$. 

In the low-$T$ phase, on the other hand, there is no spatial structure
in the diagonal noise, and no strong correlations among time-slices,
so no localisation is expected. The various TM components of the wave
function are effectively mixed by the hopping terms, and so TM 
acts effectively as a fourth dimension, i.e., one deals with
a single effectively 4D system. The larger density of low unperturbed
modes and the fact that they can easily mix is expected to lead to a
nonvanishing $\rho(0)$. 

\begin{figure}[t]
   \centering
   \includegraphics[width=0.47\textwidth]{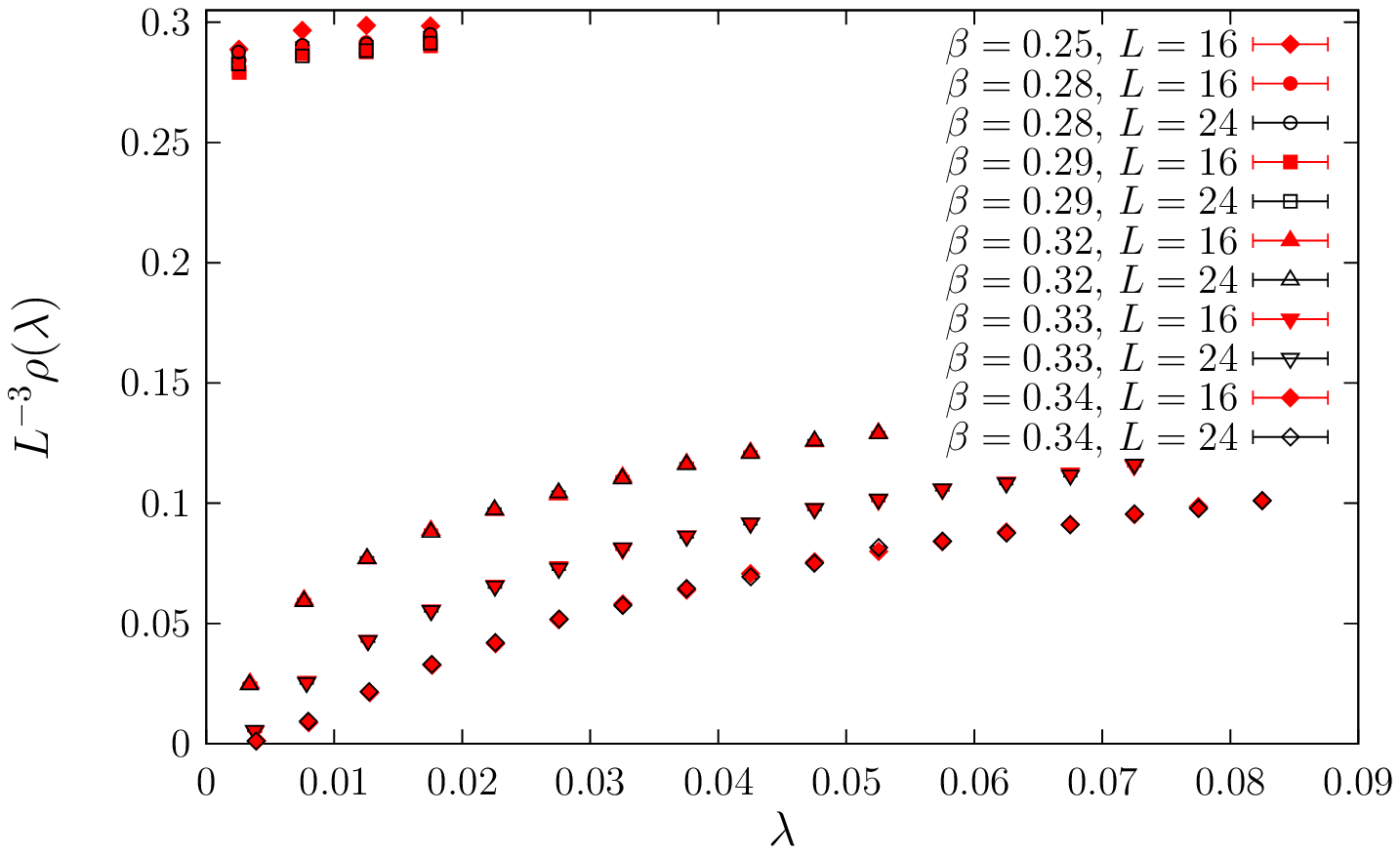}\hfil
   \includegraphics[width=0.47\textwidth]{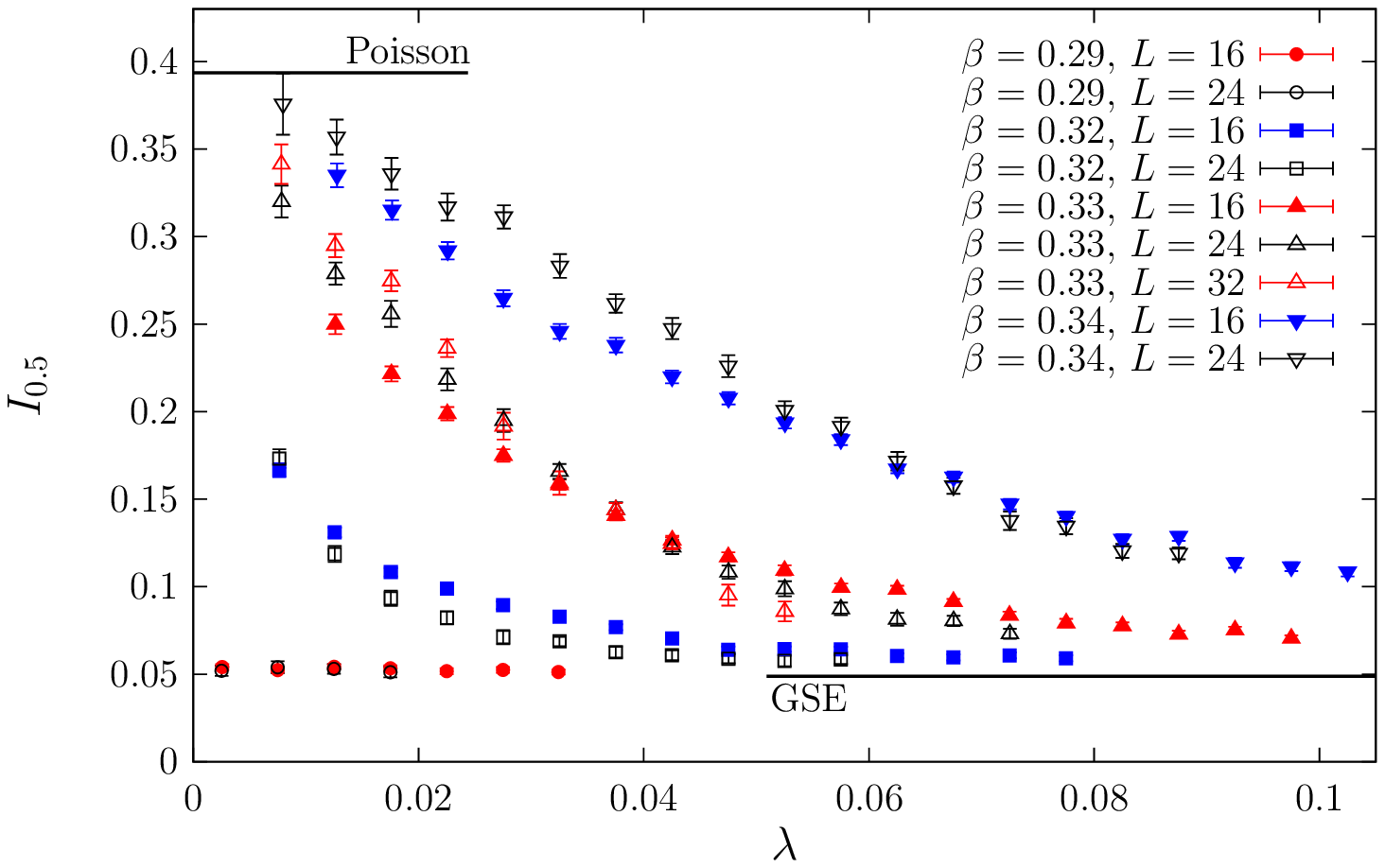}
   \caption{Spectral density $\rho$ (left) and spectral statistics $I_{0.5}$
     (right) in our toy model.}
   \label{fig:2}
 \end{figure}

Summarising, the relevant feature both for chiral symmetry restoration
and localisation of the lowest modes is expected to be the ordering of
the PLs. In the absence of ordering we expect instead to have
S$\chi$SB and delocalisation. This ultimately links both phenomena to 
the confining properties of the theory, thus providing an explanation
for the coincidence of the chiral transition, the onset of
localisation, and deconfinement, the latter being the ``fundamental''
phenomenon on which the other two depend.

\section{Toy model}
\label{sec:tm}

In order to check the arguments of the previous Section, we have
constructed a QCD-based toy model, keeping only the features that are
expected to be relevant for chiral symmetry breaking/restoration and
delocalisation/localisation of the lowest modes~\cite{GKP2}. As we
discussed above, these should be the ordering of the PLs, and the
correlation between spatial links on different time slices that this
induces. We thus considered a Hamiltonian with the same structure as
the Dirac-Anderson Hamiltonian, Eq.~\eqref{eq:d_and}, and replaced the 
PL phases $\phi_a(\vec x)$ and the spatial gauge links $\tdU_{\pm
  j}(t,\vec x)$ with appropriate toy-model counterparts $\phi_{\vec
  x}^a$ and $u_{\pm j}(t,\vec x)$. In particular, $\phi_{\vec x}^a$
are taken to be the phases of a set of complex-spin variables,
constrained by $\sum_a \phi_{\vec x}^a =0$, and $u_{j}(t,\vec x)$ are
SU$(N_c)$ matrices. Writing the Dirac-Anderson Hamiltonian as a
functional $H={\cal H}[\phi_a(\vec x),\tdU_{\pm j}(t,\vec x)]$, the
toy model Hamiltonian is then defined by $H^{\rm toy}= {\cal
  H}[\phi^a_{\vec x},u_{\pm j}(t,\vec x)]$. 

The dynamics obeyed by the toy-model variables is constructed
mimicking that of their counterparts in gauge theory. The important
feature of the PL dynamics is the existence of an ordered and of a
disordered phase. This is easily achieved for our complex spins by
taking a spin model of the following form:
\begin{equation}
  \label{eq:tm_ham_diag_dyn}
\textstyle  \beta H_{\rm noise} =  
  -\f{\beta}{N_c} \sum_{\vec x,j,a}  \cos(\phi^a_{\vec
    x + \hat\jmath}-\phi^a_{\vec x})
  - \f{2h}{N_c(N_c-1)}\sum_{\vec x,a<b}\cos(\phi^a_{\vec
    x}-\phi^b_{\vec x}) \,.
\end{equation}
For $h\neq 0$ this model has a $\mathbb{Z}_{N_c}$ symmetry, which is
spontaneously broken at large $\beta$. The system can then select one
of the $N_c$ vacua $\phi_{\vec x}^a=\f{2\pi}{N_c}$ $\forall a,\vec x$
(corresponding to the center elements of SU$(N_c)$ along which the PL
can align). The important feature of the dynamics of spatial gauge
links is instead the appearance of correlations among time-slices when
the PLs get ordered. This is achieved by using the Wilson action
without spatial plaquettes, and dropping the fermion determinant. The
effect of the PLs on the links is reproduced by coupling the toy-model 
links to the complex spin phases, treated as an external field. 
Denoting with $p(\vec x) =\diag(e^{i\phi_{\vec x}^a})$, we thus define
the toy-link action
\begin{equation}
\textstyle  S_u = -\hat\beta\,\Re\,\tr \sum_{\vec x}\sum_{j=1}^3\left\{
   u_{j}^\dag(0,\vec x)p^\dag(\vec x)u_{j}(N_T-1,\vec x)p(\vec
   x+\hat \jmath) 
+\sum_{t=0}^{N_T-2}u_{j}(t,\vec x)
  u_{j}^\dag(t+1,\vec x)
\right\}\,,
\end{equation}
where $\hat\beta$ plays the role of gauge coupling, and the
expectation value of an observable $ {\cal O}$ as 
\begin{equation}
  \label{eq:exp_val}
\textstyle  \la {\cal O} \ra = \left[\int D\phi \,e^{-\beta H_{\rm
        noise}[\phi]}\right]^{-1}
{\displaystyle\int} D\phi \,e^{-\beta H_{\rm noise}[\phi]} \left[\f{\int
        Du \, e^{-S_u[\phi,u]}{\cal O}[\phi,u]}{\int Du\,
        e^{-S_u[\phi,u]}}\right]
\,.
\end{equation}

 \begin{figure}[t]
   \centering
   \includegraphics[width=0.4\textwidth]{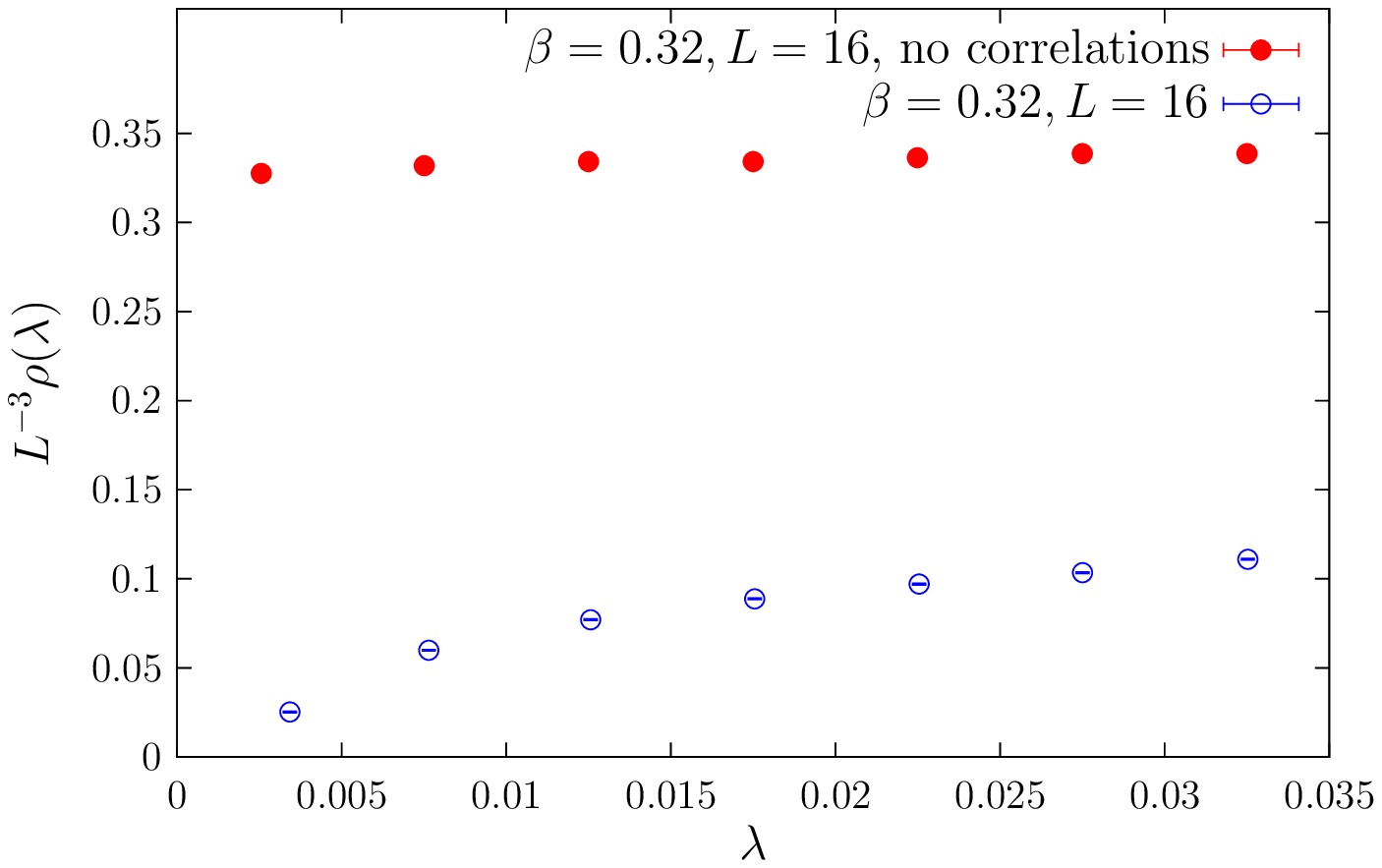}\hfil
   \includegraphics[width=0.4\textwidth]{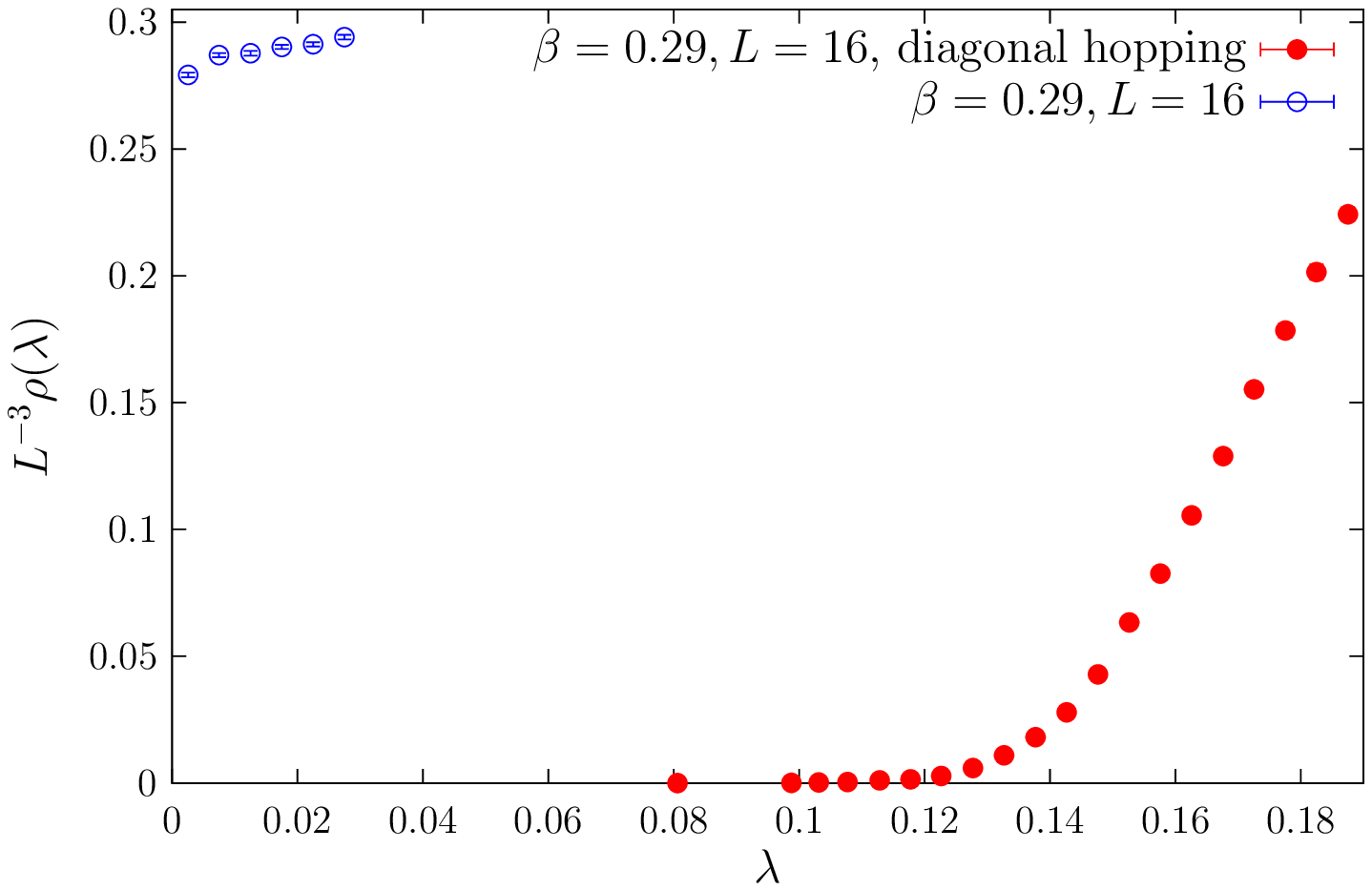}
   \includegraphics[width=0.4\textwidth]{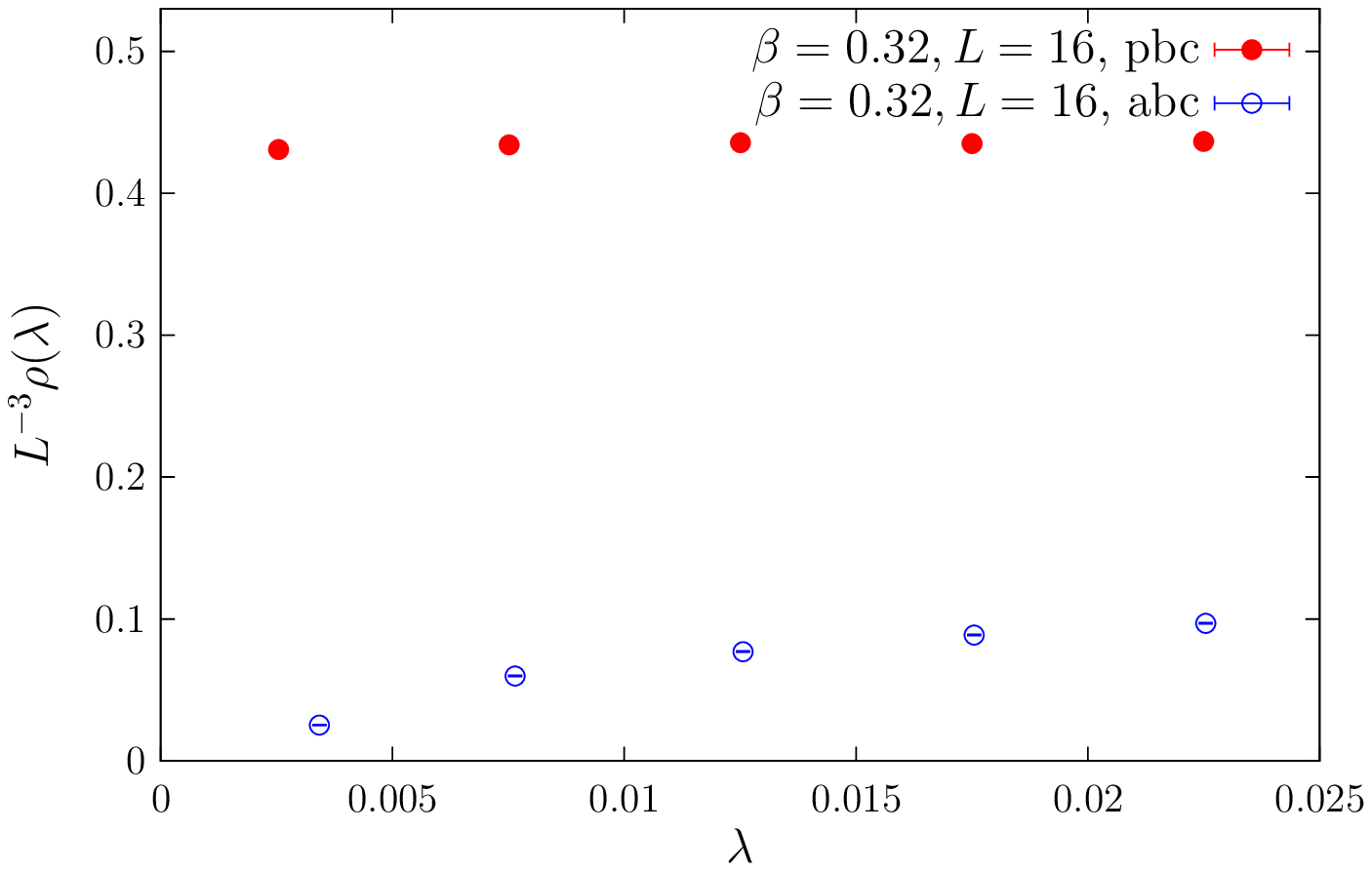}\hfil
   \includegraphics[width=0.4\textwidth]{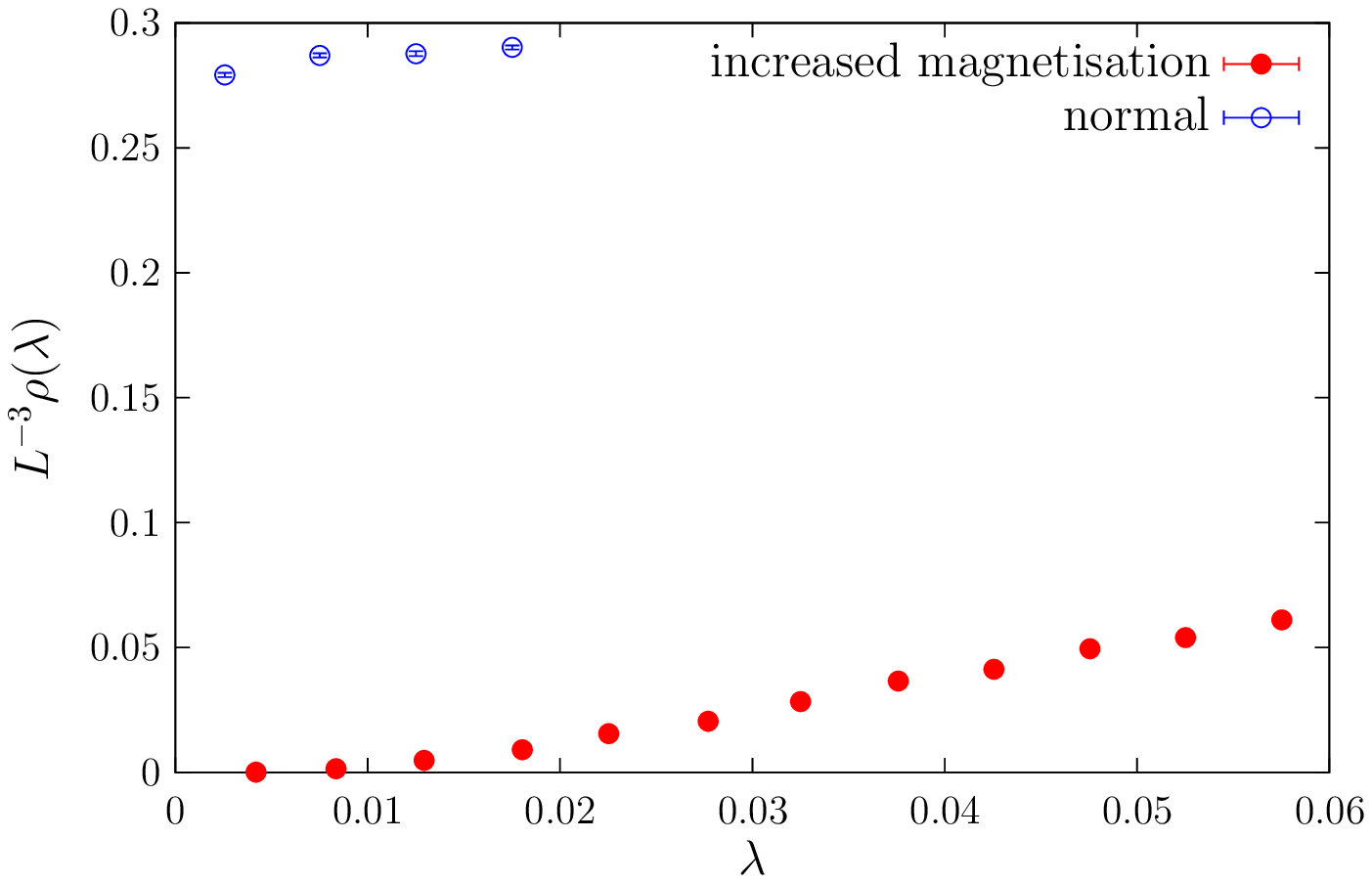}
   \caption{Spectral density in tweaked versions of the toy model:
     $\hat\beta =0$ (top left), TM-diagonal hopping terms (top right),
     temporal PBC (bottom left), and increased unperturbed
     eigenvalues (bottom right).}
   \label{fig:3}
 \end{figure}

\noindent For our numerical investigation we have employed the minimal version
of this toy model with only two colours and
two time-slices, $N_c=N_T=2$. In this case there is a single relevant
phase, $\phi_{\vec x}=\phi^1_{\vec x}=-\phi^2_{\vec x}$, and only one
relevant Matsubara frequency $\omega(\vec x)= \f{\phi_{\vec
    x}+\pi}{2}$, and the ``unperturbed'' eigenvalues are simply $\pm
\eta_4(\vec x) \cos\f{\phi_{\vec x}}{2}$. Since we are interested
mainly in the dependence on $\beta$, i.e., on the ordering of the spin
system, we fixed $h=1.0$, which leads to a critical $\beta_c\approx 0.3$, 
separating the disordered ($\beta<\beta_c$) and ordered
($\beta>\beta_c$) phases, corresponding to the confined and deconfined
phases of a gauge theory. The ``gauge coupling'' was also fixed to
$\hat \beta=5.0$. We then studied the localisation properties of the
low modes, and ``chiral symmetry breaking'', defined here as the
presence of a nonzero spectral density at the origin. 

In Fig.~\ref{fig:2}, left panel, we show the spectral density at the
low end of the spectrum, both in the disordered and in the ordered
phases. While in the disordered phase one has a nonvanishing $\rho(0)$
and so ``chiral symmetry breaking'', in the ordered phase $\rho(0)=0$
and ``chiral symmetry'' is restored. In order to detect localisation
of the eigenmodes we have exploited the fact that delocalised modes
obey RMT statistics, while localised modes obey Poisson statistics. 
We have then computed the integrated probability density $I_{0.5}$ of
the unfolded level spacings $s_i$, $I_{0.5}=\int_0^{0.5}ds\,P_\lambda(s)$, 
with $s_i=\f{\lambda_{i+1}-\lambda_i}{\la \lambda_{i+1}-\lambda_i
  \ra_\lambda}$. Here $P_\lambda(s)$ is the probability density of $s$
computed locally in the spectrum in a small interval around
$\lambda$. Our results are shown in Fig.~\ref{fig:2}, right panel,
together with the predictions of Poisson statistics and of the
appropriate ensemble of RMT, which in the case at hand is the Gaussian
Symplectic Ensemble. The transition from Poisson to RMT statistics 
is evident, and becomes sharper as the volume of the system is
increased, thus signaling the presence of a phase transition in the
thermodynamic limit. The mobility edge is determined by the crossing
point of the curves corresponding to different volumes, and increases
as the ordering of the spins increases. 

Our minimal toy model is thus able to reproduce the qualitative
features of QCD regarding chiral symmetry and localisation. By
tweaking the toy model one can further check the relative importance
of the two effects caused by ordering, i.e., the decreased density of
low unperturbed modes and the decreased mixing of the wave function's
TM components. The same consequences are expected to be observed in
QCD. In Fig.~\ref{fig:3} we show the spectral density obtained in
several modifications of the toy model. In the top left panel,
correlations across time-slices are removed by setting $\hat\beta=0$,
in the ordered phase of the spin model: as a result, chiral symmetry
gets broken. In the top right panel, the mixing between TM components
is removed by making the hopping terms diagonal in TM, in the
disordered phase: this restores chiral symmetry. In the bottom left
panel, boundary conditions in the temporal direction are switched from
antiperiodic to periodic, in the ordered phase, so increasing the
density of low unperturbed modes: chiral symmetry is broken. Finally,
in the bottom right panel we artificially increase the unperturbed
eigenvalues, in the disordered phase: chiral symmetry is restored. In
all cases, chiral symmetry restoration/breaking is accompanied by
localisation/delocalisation of the low modes. In conclusion, both
mixing of different TM components of the quark wave function (or lack
thereof) and the density of low unperturbed modes are crucial for the
fate of chiral symmetry and localisation.

\section{Conclusions}
\label{sec:concl}

Recasting the staggered Dirac operator in the form of an Anderson-like
Hamiltonian clarifies the relation between the
localisation/delocalisation transitions in QCD and in the unitary 
Anderson model. Moreover, it allows a better understanding of the
localisation mechanism at high temperature, and sheds some light on
mode delocalisation and on the formation of a chiral condensate via
accumulation of small eigenmodes at low temperature. The fate of
localisation and chiral symmetry seems to be determined by the
ordering of the Polyakov lines, and thus ultimately by the confining
properties of the theory. Further work is certainly needed along these
lines.

\end{document}